\documentclass[conference]{IEEEtran}

\usepackage[utf8]{inputenc}
\usepackage{graphicx}
\usepackage{amsmath}
\usepackage{cite}
\usepackage{url}
\usepackage[table]{xcolor}
\usepackage{tikz}
\usepackage{etoolbox} 
\usepackage{amsmath}
\usepackage{amsthm}
\usepackage[utf8]{inputenc}
\usepackage{amsmath, amssymb}
\usepackage{listings}
\usepackage{xcolor}
\usepackage{tcolorbox}

 \usepackage{booktabs}
\definecolor{keywordblue}{RGB}{0,0,139}
\definecolor{commentgray}{RGB}{96,96,96}
\definecolor{linenumgray}{RGB}{128,128,128}
\definecolor{framegray}{RGB}{192,192,192}
\lstdefinestyle{algorithmic}{
    basicstyle=\ttfamily\footnotesize,
    keywordstyle=\bfseries\color{keywordblue},
    commentstyle=\rmfamily\itshape\color{commentgray}\scriptsize,
    numberstyle=\tiny\color{linenumgray},
    identifierstyle=\color{black},
    lineskip=1pt,
    basewidth=0.48em,
    numbers=left,
    numbersep=8pt,
    stepnumber=1,
    firstnumber=1,
    showstringspaces=false,
    breaklines=true,
    breakatwhitespace=true,
    breakindent=12pt,
    frame=single,
    frameround=ffff,
    rulecolor=\color{framegray},
    framesep=4pt,
    xleftmargin=10pt,
    xrightmargin=4pt,
    tabsize=2,
    showtabs=false,
    captionpos=b,
    abovecaptionskip=6pt,
    belowcaptionskip=3pt,
    escapeinside={(*@}{@*)},
    morekeywords={Algorithm, Input, Output, Require, Ensure, for, to, do, if, then, else, end, while, repeat, until, return, begin, procedure, function},
    emphstyle=\bfseries
}

\usetikzlibrary{arrows.meta, positioning, shapes.multipart}

\title{Privacy-Preserving Anonymization of System and Network Event Logs Using Salt-Based Hashing and Temporal Noise}
\author{
    \IEEEauthorblockN{Shreyas Bargale}
    \IEEEauthorblockA{Department of Computer Science\\
    Indian Institute of Technology Madras\\
    Email: cs22b016@smail.iitm.ac.in}
    \and
    \IEEEauthorblockN{Akshit Vakati Venkata}
    \IEEEauthorblockA{Department of Metallurgical \\
    and Materials Engineering\\
    Indian Institute of Technology Madras\\
    Email: mm23b009@smail.iitm.ac.in}
    \and
    \IEEEauthorblockN{Jaimandeep Singh}
    \IEEEauthorblockA{Indian Institute of Technology Madras\\
    Email: spc701@imail.iitm.ac.in}
    \and
    \IEEEauthorblockN{Chester Rebeiro}
    \IEEEauthorblockA{Department of Computer Science\\ 
    Indian Institute of Technology Madras\\
    Email: chester@cse.iitm.ac.in}
}

\begin{document}

\maketitle

\begin{abstract}
    System and network event logs are essential for security analytics, threat detection, and operational monitoring. However, these logs often contain Personally Identifiable Information (PII), raising significant privacy concerns when shared or analyzed. A key challenge in log anonymization is balancing privacy protection with the retention of sufficient structure for meaningful analysis. Overly aggressive anonymization can destroy contextual integrity, while weak techniques risk re-identification through linkage or inference attacks. This paper introduces novel field-specific anonymization methods that address this trade-off. For IP addresses, we propose a salt-based hashing technique applied at the per-octet level, preserving both subnet and host structure to enable correlation across various log entries while ensuring non-reversibility. For port numbers, full-value hashing with range mapping maintains interpretability. We also present an order-preserving timestamp anonymization scheme using adaptive noise injection, which obfuscates exact times without disrupting event sequences. An open-source tool implementing these techniques has been released to support practical deployment and reproducible research. Evaluations using entropy metrics, collision rates, and residual leakage analysis demonstrate that the proposed approach effectively protects privacy while preserving analytical utility.

\end{abstract}

\section{Introduction}

\subsection{Motivation and Context}
System and network event logs play a pivotal role in cybersecurity, system diagnostics, compliance auditing, and behavioral analytics. These logs capture granular records of interactions between users, applications, and infrastructure components, often containing Personally Identifiable Information (PII) such as IP addresses, timestamps, usernames, and port numbers~\cite{science2024, bindplane2024}. While the analytical value of such logs is high, their use is severely constrained by privacy concerns and regulatory obligations such as the GDPR, CCPA, HIPAA, and India’s Digital Personal Data Protection Act~\cite{gdpr28,dpdp2023,ccpa2024,hipaa}.

Anonymization serves as a crucial step in making these logs shareable and analyzable without compromising user or organizational privacy~\cite{ohm2024}. However, conventional anonymization methods often present a trade-off between privacy and utility. Techniques that aggressively generalize or redact sensitive fields may preserve privacy but render the data analytically useless. Conversely, lightweight pseudonymization methods may retain analytical richness but expose the data to re-identification attacks through linkage, inference, or fingerprinting~\cite{slagell2005}.

\subsection{Challenges in Existing Log Anonymization Techniques}
While anonymization is central to privacy-preserving log sharing, current techniques often fall short in practice due to two critical issues: vulnerability to re-identification attacks and loss of analytical utility~\cite{aleroud2021}.

Many existing techniques in Table~\ref{table:field-techniques}, \ref{table:tool-comparison} rely on basic pseudonymization, prefix-preserving encryption, or random replacement strategies~\cite{flaim2006}. However, such methods can be susceptible to various types of attacks. One common threat is \textbf{dictionary and reverse lookup attacks}, especially when salting is not applied or the same salt is reused across multiple datasets. Additionally, \textbf{fingerprinting and inference attacks} may occur when consistent patterns in fields such as IP addresses, port numbers, or timestamps are preserved, allowing adversaries to infer user identities or behaviors~\cite{science2024}. Another major concern is \textbf{linkage attacks}, where anonymized data can be correlated with external datasets to re-identify individuals or systems~\cite{ohm2024}.

Moreover, overly randomized approaches—such as uncontrolled noise injection or complete redaction—can destroy temporal or structural context essential for meaningful analysis~\cite{condensation2024}. For example, full IP randomization inhibits detection of repeated access attempts, and naive timestamp obfuscation breaks event sequencing.

\begin{figure}[htbp]
\centering
\begin{tikzpicture}[
    font=\scriptsize,
    box/.style={
        rectangle, draw, rounded corners, align=left,
        text width=5.4cm, minimum height=1.1cm, inner sep=6pt
    },
    level/.style={text width=2.5cm, align=left},
    arrow/.style={thick, -{Stealth[length=2mm]}}
]
\draw[arrow] (0,4.5) -- (0,.5);
\node[rotate=90, anchor=south] at (-0.1,2.25) {\textbf{Privacy Level}};

\node[box, fill=blue!15, anchor=west] at (0.2,4.0)
{\textbf{Extreme Randomization / Generalization}\\(e.g., full redaction, heavy noise)};
\node[level, anchor=west] at (6.1,4.0)
{\textbf{Low Utility}\\Weak correlation,\\unusable for forensics};

\node[box, fill=yellow!20, anchor=west] at (0.2,2.5)
{\textbf{Field-Aware Anonymization (Proposed)}};
\node[level, anchor=west] at (6.1,2.5)
{\textbf{Moderate Utility}\\Preserves structure,\\ensures unlinkability};

\node[box, fill=red!15, anchor=west] at (0.2,1.0)
{\textbf{Weak Hashing / Pseudonymization without salt}};
\node[level, anchor=west] at (6.1,1.0)
{\textbf{High Utility}\\High risk of\\re-identification};
\end{tikzpicture}

\caption{Privacy–utility spectrum for anonymization techniques}
\label{fig:privacy_utility_spectrum}
\end{figure}
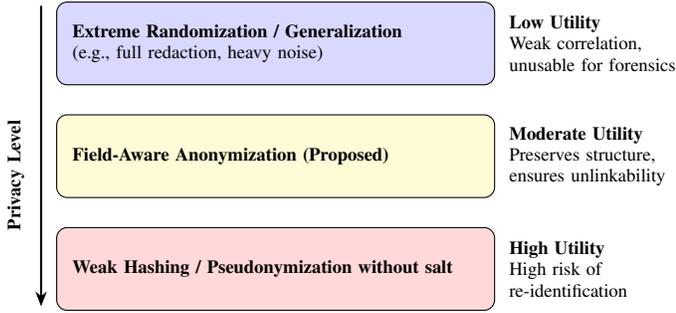

The key challenge is thus to strike a balance between privacy and analytical value as shown in Fig.~\ref{fig:privacy_utility_spectrum}. Excessive obfuscation undermines security monitoring, while insufficient protection risks privacy breaches. This paper aims to address this challenge through structured, field-aware anonymization.

\subsection{Our Contribution}
This paper proposes a set of novel anonymization techniques specifically designed for field-aware protection of system and network event logs. Key contributions include:
\begin{itemize}
    \item A \textbf{salt-based hashing mechanism} for IP anonymization applied at the octet level, preserving both subnet and host structure to support cross-entry correlation within a log file.
    \item A \textbf{temporal anonymization technique} using adaptive noise injection to retain event order while masking precise timestamps.
    \item An \textbf{open-source log anonymization tool} released on GitHub, supporting multiple log formats and customizable field-specific transformations.
    \item A \textbf{comprehensive evaluation framework} based on entropy, collision analysis, and residual leakage to assess privacy-utility trade-offs.
\end{itemize}

\subsection{Structure of This Paper}
The remainder of this paper is organized as follows: Section~\ref{sec:related} reviews existing anonymization techniques and their limitations. Section~\ref{sec:methods} introduces the field-specific methods used in our framework. Section~\ref{sec:timestamp} describes the adaptive noise-based timestamp anonymization scheme.Section~\ref{sec:iphashing} presents the salt-based IP anonymization approach and its security properties. Section~\ref{sec:security} provides experimental results and analysis.Section~\ref{sec:tool} discusses the implementation of the open-source tool and its integration with real-world logs. Finally, Section~\ref{sec:conclusion} concludes the paper and outlines future work.

\section{Background and Review of Log Anonymization Techniques}
\label{sec:related}
Anonymization involves removal/modifying of data to prevent re-identification. In our use case we need the anonymization techniques to make it impossible to trace the logs back to the system which generated it at the same time retain enough information and structure for analysis.

\subsubsection{Review of key Anonymization Techniques}

\begin{itemize}
    \item \textbf{Prefix-Preserving Encryption (PPE):} This involves anonymizing IP addresses while preserving their subnet relationships, making it useful for subnet based analysis. A similar approach is \textbf{format-preserving anonymization}, which maintains the structure of data but alters the values \cite{xie2021generalized,fan2002prefix}.

    \item \textbf{Differential Privacy:} This involves adding  random noise based on some function to data values or data clusters. This method ensures that individual entries do not significantly affect the output. It can be combined with k-means clustering and PPE to increase anonymity and structure. Can be used for timestamps \cite{nissim2017dp,aleroud2021}.

    \item \textbf{Truncation and Filtering:} This involves generalizing or completely removing specific fields. Eg. \textit{Black Marker} technique, which replaces sensitive values like IP addresses with fixed constants (e.g., 0.0.0.0).

    \item \textbf{Random Permutation and Replacement:} These techniques permute or randomly replace values in fields to hide patterns. For textual data, predefined terms may be substituted with arbitrary strings, which can be of same length to preserve format. Each unique information can be mapped to a specified format enumerated data and then the mapping can be hidden.

    \item \textbf{Enumeration:} This method is particularly useful in scenarios where maintaining the original order of data is important, but the actual values must be concealed. It works by replacing sensitive information with sequential identifiers, thereby preserving the sequence while ensuring privacy. Additionally, placeholders can be employed to obscure metadata without disrupting the structural integrity of the data.

    \item \textbf{Hashing (with Salting):} Cryptographic hash functions (e.g., SHA-256, MD5) can anonymize fields irreversibly. Adding a secret salt increases resistance to brute-force or dictionary attacks. This provided inspiration for the techniques we will recommend.
    This salt can be randomised and deleted to make things irreversible but deterministic after salt generation.
    \item 
    \textbf{Time shift}:
    Adding a random shift across all timestamps to prevent use of publicly known information to backtrace.
\end{itemize}

\subsection{Field-Specific Anonymization Strategies and Tools}

An effective anonymization pipeline requires field-specific strategies, particularly for network log attributes considered personally identifiable information (PII). The key PII elements identified include IP addresses, MAC addresses, timestamps, payloads (in certain cases), port numbers, and TCP sequence numbers. Table~\ref{table:field-techniques} outlines suitable anonymization techniques and tools for these fields \cite{Aghili2024-mj}. 

While tools such as \textit{Logstash} and \textit{LogLicker} offer built-in support for log anonymization, they often require substantial configuration and may not provide the flexibility needed for fine-grained control. Custom scripting, on the other hand, allows for more adaptable and context-sensitive anonymization.

 To provide further context, Table~\ref{table:tool-timeline} presents a timeline of key anonymization tools, and Table~\ref{table:tool-comparison} compares their flexibility and privacy capabilities\cite{dijkhuizen2018survey}.

\begin{table}[htbp]
    \centering
    \caption{Field-Specific Techniques and Tools}
    \label{table:field-techniques}
    \renewcommand{\arraystretch}{1.5}
    \resizebox{\columnwidth}{!}{%
    \begin{tabular}{|l|l|l|}
        \hline
        \textbf{Field} & \textbf{Techniques} & \textbf{Tools} \\\hline
        IP Address & PPE, random permutation, clustering + hashing & CryptoPan, TCPDpriv, FLAIM \\\hline
        MAC Address & Hashing, vendor-prefix preservation & FLAIM, custom scripts \\\hline
        Timestamps & Noise injection, rounding, randomisation & Python, PCAPLib \\\hline
        Payload & Suppression, tokenization, random substitution & Wireshark, PCAPLib \\\hline
        Port Numbers & Classify, substitute, reuse of IP techniques & Scapy, FLAIM \\\hline
        TCP Seq. Numbers & Permutation, hashing, Randomisation & CryptoPan, custom tools \\\hline
    \end{tabular}
    }
\end{table}

\begin{table}[htbp]
\centering
\caption{Timeline of Key Anonymization Tools}

\label{table:tool-timeline}
\renewcommand{\arraystretch}{1.5}
    \resizebox{\columnwidth}{!}{%
\begin{tabular}{|l|p{10cm}|}
\hline
\textbf{Tool} & \textbf{Description} \\
\hline
TCPdpriv & Early tool using prefix-preserving anonymization; relies on static lookup tables which limit scalability. \\
Crypto-PAn & AES-based prefix-preserving encryption; supports consistent anonymization and is memory efficient. \\
PktAnon & Supports parsing and replacement based on protocol in packet \\
PCAPLib & Uses Wireshark dissectors to anonymize application-layer fields. \\
TraceWrangler & GUI-based tool supporting protocol tunneling and sanitization of PCAP/PCAPng files. \\
LogLicker & Tool for system/application log anonymization using rule-based(regex) redaction and hashing. \\
Logstash & Modern, flexible pipeline tool with plugin and scripting support for anonymization. \\
Custom Script & Custom anonymization logic using salted hashes, prefix-preserving transformations, and differential privacy. \\
\hline
\end{tabular}
}
\end{table}

\begin{table}[htbp]
\centering
\caption{Feature Comparison of Anonymization Tools}
\renewcommand{\arraystretch}{1.5}
    \resizebox{\columnwidth}{!}{%
\label{table:tool-comparison}
\begin{tabular}{|l|c|c|c|c|p{5.5cm}|}
\hline
\textbf{Tool} & \textbf{Flexibility} & \textbf{Privacy Features} \\
\hline
TCPdpriv & Low & Prefix-preserving only \\
Crypto-PAn & Moderate & AES-based prefix-preserving encryption \\
PktAnon & High & Modular, protocol-specific anonymization \\
PCAPLib &  Moderate & Application-layer awareness using dissectors \\
TraceWrangler  & High & GUI-based tool supporting protocol tunneling \\
LogLicker & Low & Rule-based redaction and hashing \\
Logstash & Very High & Plugin-based scripting for flexible anonymization \\
Custom Script  & High & Salted hashing, prefix-preserving, differential privacy \\
\hline
\end{tabular}
}
\end{table}

\subsection{Overview of ARX Framework and Tool}
ARX is a data anonymization framework designed to address the challenges of protecting sensitive information in datasets. Its architecture and feature set enable it to implement a variety of privacy models, ensuring a balance between privacy protection and data utility\cite{prasser2014arx,prasser2020flexible}.

Its architecture supports multiple privacy models including $\delta$-presence, $\beta$-likeness, differential privacy, and ($\kappa$,$\epsilon$)-anonymity \cite{holohan2017kanonymity}, all of them have mathematically grounded proofs which guarantees protection against different disclosure risks. This allows users to implement privacy models to protect their data and also comply with regulatory bodies.

For network log anonymization, ARX tool can be used with specialized techniques like prefix-preserving IP anonymization, temporal degradation of timestamps, and consistent pseudonymization of entity identifiers. By implementing privacy models like k-anonymity\cite{sweeney2002kanonymity} to make each field indistinguishable from others, differential privacy to protect against inference attacks, and t-closeness\cite{li2007tcloseness} to protect sensitive attributes, organizations can effectively anonymize network logs while preserving sufficient information for meaningful analysis of traffic patterns, attack signatures, and temporal activity.

\subsubsection{Supported Privacy Models}

ARX brings together a range of privacy models, each designed to tackle different types of disclosure risks. The $\delta$‑presence model ensures that the likelihood of any individual being part of the anonymized dataset stays within a defined range—between $\delta_{\min}$ and $\delta_{\max}$—though it requires access to the full population and can be computationally demanding. The $\beta$‑likeness model controls how much the probability of a sensitive value can shift between the overall data and any anonymized group, offering an intuitive way to assess disclosure risk, though sometimes requiring broad generalization when dealing with outliers or multiple sensitive attributes. Differential privacy, supported in both its strict $\epsilon$ form and the relaxed $(\epsilon,\delta)$ version, protects individual records by adding carefully calibrated noise, usually via the Laplace mechanism. This approach provides strong theoretical guarantees, but choosing the right $\epsilon$ remains critical to maintaining a balance between privacy and data usefulness. Lastly, the $(k,\epsilon)$‑anonymity model combines the strengths of $k$‑anonymity and differential privacy by ensuring each group has at least $k$ records while also bounding the likelihood of sensitive value disclosure, thus offering both group-level anonymity and individual-level protection.

\subsubsection{Limitations}
While ARX offers many advantages, it also comes with some challenges. To use the $\delta$‑presence model, users need access to the full population dataset, which is often not available in real-world situations. Some techniques—such as running checks for $\delta$‑presence or fine-tuning $(k,\epsilon)$‑anonymity—can also be \emph{computationally expensive}, especially when dealing with large or complex datasets. Another key issue is the \emph{trade-off between privacy and data usefulness}: tighter privacy settings, like smaller values of $\delta_{\text{max}}$, $\epsilon$, or $\beta$, usually lead to more data being generalized or removed, which can reduce its value for analysis. Lastly, picking the right settings isn’t always straightforward. It requires a good understanding of the data and the specific goals of the project, making it hard to define one-size-fits-all recommendations.
\section {Existing IP anonymization algorithms}
\label{sec:methods}
Numerous anonymization algorithms have been proposed to address the challenges of preserving privacy in IP address data. These methods vary widely in complexity, effectiveness, and suitability for analytical tasks. In this section, we review some of the key existing approaches that have been explored for anonymizing IP addresses, along with their underlying principles and known limitations.
\subsection{Data Anonymization -Condensation with differential privacy and Its Analytical Limitations}

The condensation-based anonymization technique, proposed by \textit{Aggarwal and Yu }\cite{aggarwal2008condensation}, balances privacy and data utility by preserving statistical characteristics while anonymizing sensitive details. The method involves the following key steps:

\begin{enumerate}
    \item \textbf{Clustering:} Data is grouped into clusters of at least \(k\) entries to satisfy \(k\)-anonymity.
    \item \textbf{IP Address Anonymization:}
     A \textit{prefix-preserving permutation} is applied to the network portion of each IP, to maintain hierarchical structure.The remaining portion is anonymized by clustering similar IPs and replacing with the cluster mean.
    \item \textbf{Non-IP Features:} Network features such as payload size information are anonymized using \(k\)-means clustering to preserve the importance of value/quantity.
    \item \textbf{Differential Privacy:} Laplace noise is added to the above steps to increase privacy:
    \[
    \eta \sim \text{Lap}\left(\frac{\text{sensitivity}}{\epsilon}\right)
    \]
\end{enumerate}
\textbf{Analytical Limitation:} This combined approach provides strong privacy protection but makes data analysis difficult. The prefix-preserving method hides individual host identities while keeping network groups intact. However, the clustering and mean substitution steps remove differences between individual hosts within each cluster. When k-anonymity averaging is combined with differential privacy noise, the data quality becomes severely degraded. This two-part anonymization process makes statistical analysis (e.g., host-level behavior analysis, anomaly detection, and traffic pattern analysis) ineffective because the original data patterns are hidden by both the averaging from k-anonymity and the noise added by differential privacy.

\subsection{Logical Group-Based Mapping for Anonymization of IP}

Another anonymization approach maps each IP address based on its logical function in the network. It has a 3 step process which provides privacy but reduces granular IP analysis. This limits its use case given our objectives but is an interesting technique. \\
\textit{Step 1: Defining Logical Roles}, Network administrators define logical host groups (e.g., \textit{Users}, \textit{Servers}, \textit{External}). Hosts are mapped to groups using range of IP addresses. \\
\textit{Step 2: Two-Function Anonymization}, Each IP address $A$ undergoes two mappings:
\begin{itemize}
  \item \textbf{Host Identity Mapping:} A randomized pseudonym $p = h(A)$ is generated using a Random Oracle. It is irreversible and  unlinkable to the original IP.
  \item \textbf{Group Identity Mapping:} The address is also mapped to a logical group $g(A)$ for consistent anonymization within group ranges.
\end{itemize}  
\textit{Step 3: Final Address Computation}, The final anonymized IP $A'$ is computed using the group-based range and the pseudonym:
\[
A' = G[i]_0 + (p \bmod (G[i]_1 - G[i]_0))
\]
where:
\begin{itemize}
  \item $G[i]_0$ is the starting IP of the anonymized range for group $i$.
  \item $G[i]_1$ is the ending IP of the anonymized range.
  \item $p$ is the pseudonym derived from $h(A)$.
\end{itemize}
Pseudonymization ensures non reversibility and group anonymizations masks identity.  

\textit{Analysis Trade-off}: Although the mapping preserves group-level semantics, it disrupts prefix and host structure. This makes analysis such as traffic analysis, detecting subnet-specific behavior ineffective, reducing the analytical value.

\section{Anonymization Models for Network Logs}

Network logs contain detailed information about network activities including IP addresses, timestamps, user identifiers, and protocol usage patterns. This information can reveal user behavior, system vulnerabilities, and organizational network structures. Before such logs can be shared with external parties or analyzed by third-party systems, robust anonymization techniques must be applied to protect individual privacy without compromising data usefulness.

The main purpose of log anonymization is to enable analysis through Security Information and Event Management (SIEM) solutions while maintaining privacy protection. SIEM systems depend on log data to identify suspicious activities and generate security alerts, but sharing raw logs creates privacy risks.

A critical requirement for practical log anonymization is format preservation. In order to avoid custom decoders for new formats of logs, it is convenient to maintain the original format of logs\cite{rasic2023}. This ensures that existing SIEM tools can process anonymized logs without requiring additional configuration. The challenge lies in balancing privacy protection, data utility, and format compatibility without changing log structure or removing essential fields.

To address these challenges, we propose two anonymization methods: Order-Preserving Timestamp Anonymization with Adaptive Noise and IP and Port Anonymization via Salt-Based Hashing.

\subsection{\textbf{Order-Preserving Timestamp Anonymization with Adaptive Noise}}
\label{sec:timestamp}
\begin{figure}[ht]
    \centering
    \includegraphics[width=1\linewidth]{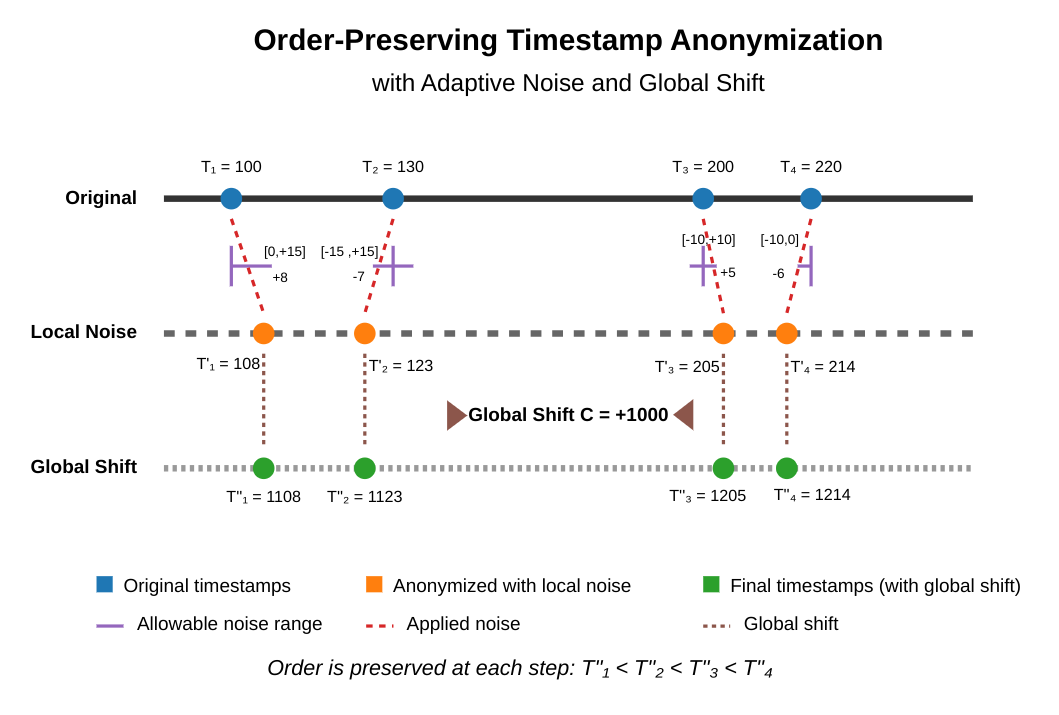}
    \caption{Order-Preserving Timestamp Anonymization with Adaptive Noise}
    \label{fig:IP Anonymisation}
\end{figure}

Temporal information in log data presents unique anonymization challenges as timestamps often contain sensitive patterns that can reveal user behavior, system usage patterns, or operational schedules. Traditional anonymization approaches either completely remove temporal data or apply uniform noise that can disrupt chronological ordering, making subsequent analysis ineffective. The preservation of event sequence is critical for many analytical tasks including anomaly detection, workflow analysis, and performance monitoring. Therefore, an anonymization technique that maintains chronological integrity while protecting temporal privacy is essential for practical log anonymization systems. To anonymize timestamps  we propose an adaptive noise-based approach. Unlike fixed-range noise, this method adjusts the allowable perturbation for each timestamp based on the distance to its neighboring events, ensuring that the relative order is never violated. This method is explained in detail in Fig.~\ref{fig:IP Anonymisation}.The process begins by sorting all events chronologically. For each timestamp $T_i$ (where $1 < i < n$), we calculate a maximum allowable noise range as follows:

\[
\Delta_i = \min\left(\frac{T_{i+1} - T_i}{2}, \frac{T_i - T_{i-1}}{2}\right)
\]

A random noise value $\delta_i$ is then sampled uniformly from the interval $[-\Delta_i, +\Delta_i]$, and the timestamp is perturbed accordingly:

\[
T_i' = T_i + \delta_i
\]

For edge cases, such as the first and last timestamps, the range is computed using only the adjacent value:
\begin{align*}
\delta_1 &\in \left[0, \frac{T_2 - T_1}{2}\right] \\
\delta_n &\in \left[-\frac{T_n - T_{n-1}}{2}, 0\right]
\end{align*}

This approach guarantees that the anonymized timestamps $T_i'$ maintain their original ordering:
\[
T_1' < T_2' < \dots < T_n'
\]

To further increase privacy, a constant global offset $C$ which can be randomly generated can be added to all timestamps:
\[
T_i'' = T_i' + C
\]

This additional step ensures that the anonymized log is temporally decoupled from the original event timeline, preventing correlation attacks.

This adaptive approach achieves an optimal trade-off between data utility and privacy protection by maintaining chronological sequence integrity while preventing temporal inference attacks. The locally adaptive noise mechanism ensures that privacy protection scales with event density - sparse events receive broader anonymization ranges while densely packed sequences maintain sufficient obfuscation without violating ordering constraints. By combining order preservation with adaptive perturbation and global offset techniques, the method effectively counters both direct timestamp correlation attacks and indirect inference through inter-event timing analysis. The technique is most effective for applications that analyze event sequences rather than exact timing, including workflow analysis, system behavior monitoring, and log-based security investigations.

\subsubsection{Security Analysis Against Temporal Attacks}

The proposed order-preserving timestamp anonymization technique provides robust protection against several categories of temporal attacks that commonly target log data systems. Backdating attacks occur when malicious actors manipulate system clocks to make activities appear as though they occurred at earlier times, often to cover up unauthorized access or illegal activities. Similarly, replay attacks \cite{song2023replay} involve adversaries intercepting legitimate timestamp tokens and reusing them for unauthorized access. The adaptive noise technique creates protection against both attacks by introducing temporal uncertainty into event timing, making it impossible to determine precise original timestamps and effectively neutralizing attempts to manipulate time-based evidence while preserving chronological sequence for legitimate analysis.

Correlation attacks exploit timing patterns to map network behaviors or identify user activities within anonymized datasets, with attackers conducting systematic probing using patterns they can later identify in captured logs. The adaptive noise approach disrupts these attempts by introducing variable perturbations that prevent matching probe patterns with log entries. The randomized noise ensures identical actions have different temporal signatures while preserving chronological order for legitimate analysis \cite{zhang2015time}, making it computationally infeasible for attackers to establish reliable correlations between their known activities and anonymized log data.

Profiling attacks attempt to link records across datasets from different time periods by establishing temporal connections between data points to build comprehensive user or system profiles. The combination of adaptive noise and global offset creates robust defense against both correlation and profiling attempts by making it computationally impractical to establish temporal bridges between datasets. The global offset component creates temporal disconnection between original and anonymized data, preventing attackers from correlating intercepted timestamps with anonymized entries, while the obfuscated timestamps eliminate time-based fingerprints that these sophisticated attacks depend upon.

\newpage
\subsection{ \textbf{IP and Port Anonymization via Salt-Based Hashing:} }
\label{sec:iphashing}
We propose a new anonymization technique specifically tailored for IP addresses and port number fields. The primary objective is to maintain \textbf{file-level consistency}, ensuring that the same input value is consistently mapped to the same anonymized value within a particular log file. This is better than just plain hashing as it is susceptible to brute force attacks. The advantage this method gives over the above mentioned methods is that it retains the relation between the IPs as each repeated IP in the file gets mapped to the same IP after anonymization. It also incorporates retention of subnet relation to some extent.

Our method uses \textbf{salt-based hashing}, where a random salt is generated once per file(can be extended to a group of files processed together). Traditional encryption or hashing techniques often produce large or irregular values that do not follow the expected format of IP addresses. To address this, we split each IP address into its four octets and apply the salt-based hash function individually to each octet.

\begin{figure}[ht]
    \centering
    \includegraphics[width=0.9\linewidth]{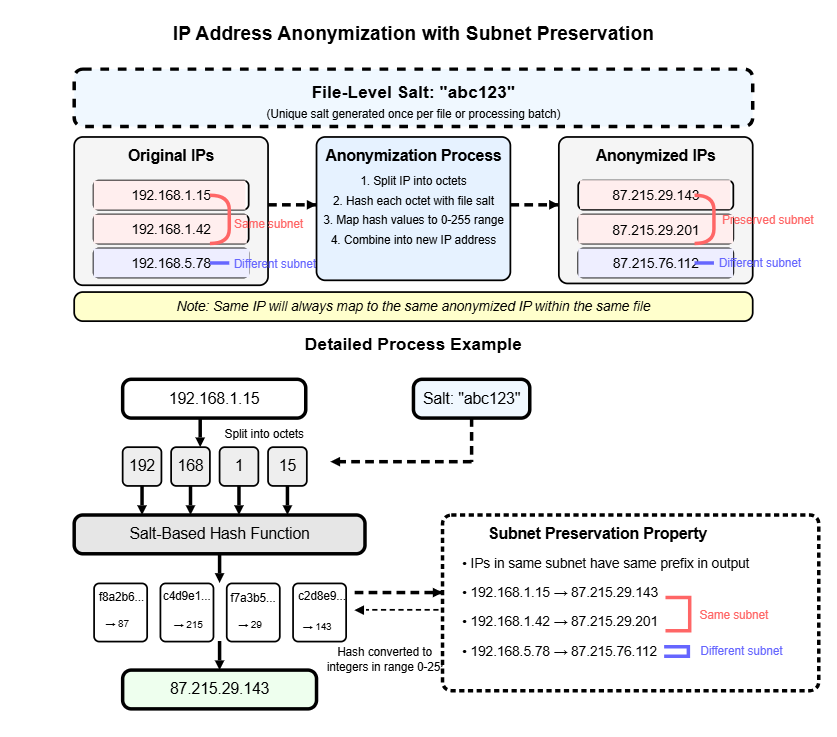}
    \caption{IP Address Anonymization with Subnet Preservation for Multiple Addresses}
    \label{fig:algohash}
\end{figure}

The resulting hash values are then mapped to integers in the range \textbf{0--255}---either through a modulus operation or via a deterministic mapping (e.g., enumeration/permutation with stored lookup). This process aims to preserve the \textbf{subnet structure to some extent} while still anonymizing the original IPs as explained in Fig.~\ref{fig:algohash}. Importantly, the same original IP will always be mapped to the same anonymized IP within the file.

A similar approach is used for port numbers, where hashed values are mapped into the range \textbf{0--65535}, preserving the valid port number space.

\subsubsection{Problem Formulation and Mathematical Foundation}

Given a dataset $\mathcal{D}$ containing potentially sensitive IPv4 addresses, we seek to construct a privacy-preserving transformation $\Phi$ that maintains analytical utility while ensuring computational anonymity. The transformation must preserve network topology characteristics, maintain consistency across identical inputs, and provide cryptographic security guarantees.

The IPv4 address space is defined as $\mathcal{A} = [0,255]^4 \cap \mathbb{Z}^4$, where each address $\boldsymbol{\alpha} \in \mathcal{A}$ is represented as a 4-tuple $\boldsymbol{\alpha} = (\alpha_1, \alpha_2, \alpha_3, \alpha_4)$ with octets $\alpha_i \in [0,255] \cap \mathbb{Z}$.
An IP dataset is a multiset $\mathcal{D} = \{\boldsymbol{\alpha}_1, \boldsymbol{\alpha}_2, \ldots, \boldsymbol{\alpha}_n\}$ where $\boldsymbol{\alpha}_i \in \mathcal{A}$ and duplicate addresses are permitted to reflect real-world network traffic patterns. Let $\mathcal{H}: \{0,1\}^* \rightarrow \{0,1\}^{256}$ denote the SHA-256 cryptographic hash function satisfying standard cryptographic properties: preimage resistance, second preimage resistance, and collision resistance. For each octet position $i \in \{1,2,3,4\}$, define a mapping function $\mathcal{M}_i: [0,255] \rightarrow [0,255]$ that deterministically transforms octets while preserving the valid IPv4 range.

The anonymization function $\Phi$ exhibits deterministic behavior for any fixed cryptographic salt $\boldsymbol{s}$ and IP dataset $\mathcal{D}$. This fundamental property ensures that identical IP addresses consistently produce identical anonymized results, formally expressed as:
\begin{equation}
\forall \boldsymbol{\alpha}_i, \boldsymbol{\alpha}_j \in \mathcal{D}: \boldsymbol{\alpha}_i = \boldsymbol{\alpha}_j \Longrightarrow \Phi(\boldsymbol{\alpha}_i) = \Phi(\boldsymbol{\alpha}_j)
\end{equation}
This consistency stems from the deterministic nature of the underlying cryptographic hash function $\mathcal{H}$, which produces constant outputs for identical input pairs $(\text{octet}, \boldsymbol{s})$. Combined with the deterministic modulo operation and position-specific mappings $\mathcal{M}_i$, this ensures reproducible anonymization results across multiple executions of the algorithm on the same dataset.

\begin{figure}[ht]
    \centering
    \includegraphics[width=0.85\linewidth]{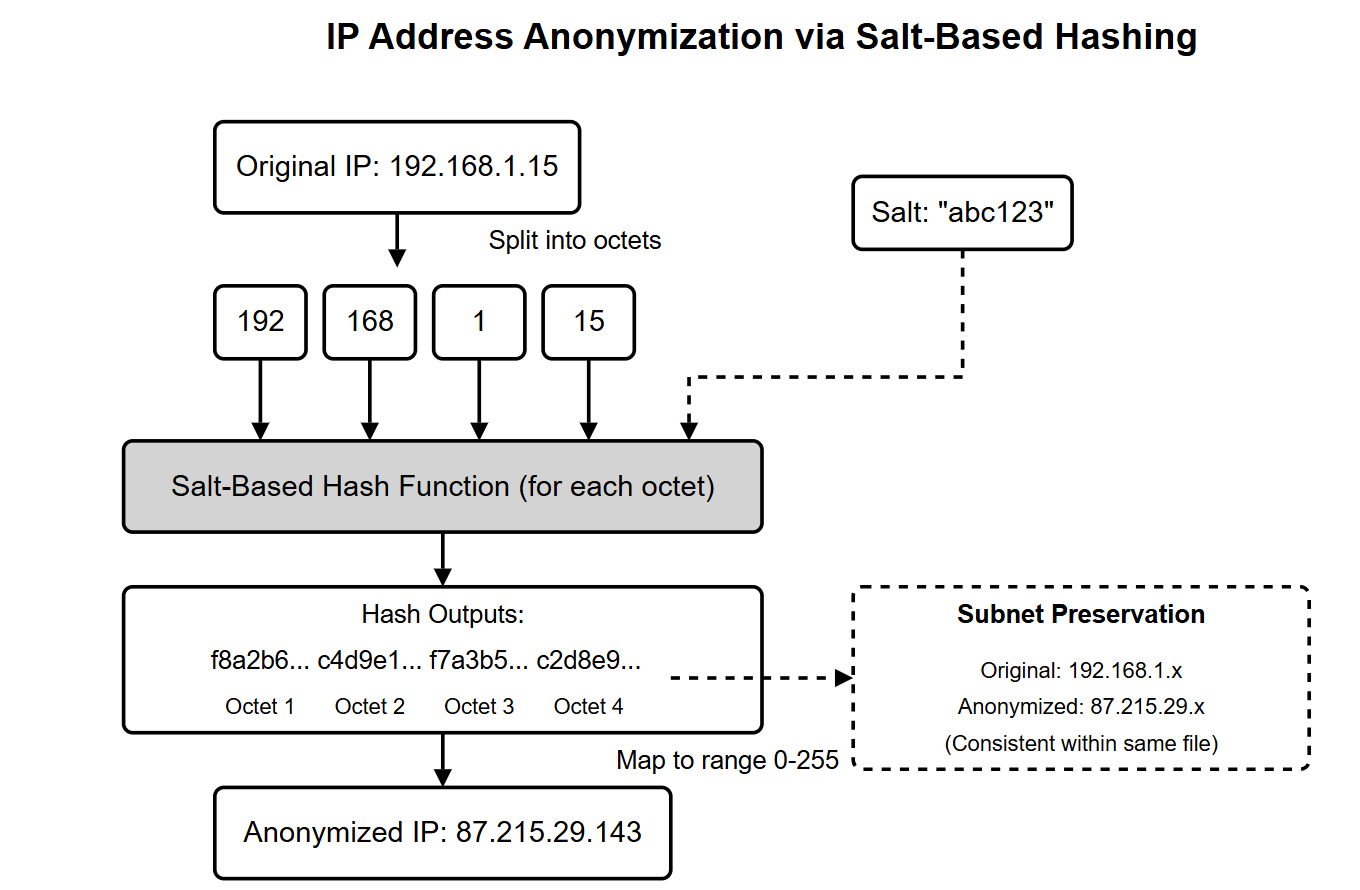}
    \caption{Control flow}
    \label{fig:IP Anonymisation 1}
\end{figure}

The algorithm also explained in Fig.~\ref{fig:IP Anonymisation 1} exhibits optimal computational efficiency with time complexity $\mathcal{O}(n + k)$ where $n = |\mathcal{D}|$ represents the total number of IP addresses in the dataset and $k = \sum_{i=1}^{4} |\mathcal{U}_i|$ denotes the aggregate number of unique octets across all positions. The algorithm processes each IP address exactly once, contributing $\mathcal{O}(n)$ to the complexity. Hash computations occur only for unique octets, bounded by $k$, with each SHA-256 computation requiring $\mathcal{O}(1)$ time for fixed-size inputs. The space complexity is $\mathcal{O}(k)$, representing the storage requirements for the four position-specific mapping tables that store the anonymized values for each unique octet at each position.

\onecolumn

\subsubsection{Algorithmic Framework}
The proposed anonymization framework operates through four distinct computational phases: \textit{initialization}, \textit{cryptographic mapping generation}, \textit{systematic anonymization}, and \textit{dataset reconstruction}.

\begin{lstlisting}[style=algorithmic, caption={Cryptographic Salt-Based IP Address Anonymization with Position-Preserving Transformations }, label={alg:crypto_ip_anonymization_alt}]
(*@\textbf{Require:}@*) IP dataset (*@$\mathcal{D} = \{\mathbf{ip}_1, \mathbf{ip}_2, \ldots, \mathbf{ip}_n\}$@*) where (*@$\mathbf{ip}_i \in \mathbb{A} \cup \mathbb{A}^{\perp}$@*)
(*@\textbf{Require:}@*) Cryptographic salt (*@$\boldsymbol{\sigma} \in \{0,1\}^*$@*)
(*@\textbf{Require:}@*) Cryptographic hash function (*@$\mathcal{H}: \{0,1\}^* \rightarrow \{0,1\}^{256}$@*)
(*@\textbf{Ensure:}@*) Anonymized dataset (*@$\mathcal{D}' = \{\mathbf{ip}'_1, \mathbf{ip}'_2, \ldots, \mathbf{ip}'_n\}$@*)

(*@\textbf{Phase I: Initialization}@*)
(*@$\mathcal{T}_j \gets \emptyset$@*) for (*@$j \in \{1,2,3,4\}$@*)    // Initialize position-specific transformation tables
(*@$\mathcal{D}' \gets \emptyset$@*)    // Initialize anonymized dataset

(*@\textbf{Phase II: Main Processing Loop}@*)
for (*@$k = 1$@*) to (*@$|\mathcal{D}|$@*) do
    (*@$\mathbf{ip}_k \gets \mathcal{D}[k]$@*)    // Extract current IP address from dataset
    if (*@$\mathbf{ip}_k \in \mathbb{A}$@*) then    // Process valid IPv4 addresses
        Parse (*@$\mathbf{ip}_k = \langle o_1, o_2, o_3, o_4 \rangle$@*) where (*@$o_i \in \mathbb{O}$@*)    // Decompose IP into octets
        for (*@$j = 1$@*) to (*@$4$@*) do    // Transform each octet with position-specific mapping
            if (*@$o_j \notin \text{dom}(\mathcal{T}_j)$@*) then    // Generate new cryptographic mapping
                (*@$\xi \gets \boldsymbol{\sigma} \parallel \text{enc}(o_j) \parallel \text{enc}(j)$@*)    // Construct salted input with position encoding
                (*@$\delta \gets \mathcal{H}(\xi)$@*)    // Apply SHA-256 cryptographic hash
                (*@$\mathcal{T}_j[o_j] \gets \text{bin2int}(\delta) \bmod 256$@*)    // Store mapping in valid octet range
            end if
            (*@$o'_j \gets \mathcal{T}_j[o_j]$@*)    // Retrieve transformed octet from mapping table
        end for
        (*@$\mathbf{ip}'_k \gets \langle o'_1, o'_2, o'_3, o'_4 \rangle$@*)    // Reconstruct anonymized IP address from octets
    else
        (*@$\mathbf{ip}'_k \gets \mathbf{ip}_k$@*)    // Preserve malformed IP addresses unchanged
    end if
    (*@$\mathcal{D}' \gets \mathcal{D}' \cup \{\mathbf{ip}'_k\}$@*)    // Append transformed IP to anonymized dataset
end for

(*@\textbf{Phase III: Finalization and Output}@*)
Verify (*@$|\mathcal{D}'| = |\mathcal{D}|$@*)    // Ensure dataset cardinality is preserved
Output (*@$\mathcal{D}'$@*)    // Complete anonymized dataset with preserved structural relationships
\end{lstlisting}

\vspace{-5pt}
\subsubsection{Notation Summary}
{\footnotesize
\begin{center}
\begin{tabular}{@{}p{0.4\linewidth}p{0.53\linewidth}@{}}
\textbf{Description} & \textbf{Symbol} \\
\midrule
Valid IPv4 address space & $\mathbb{A} = \{(o_1,o_2,o_3,o_4) : o_i \in [0,255] \cap \mathbb{Z}\}$ \\
Malformed IP address space & $\mathbb{A}^{\perp} = \{\text{invalid IP representations}\}$ \\
Octet domain & $\mathbb{O} = [0,255] \cap \mathbb{Z}$ \\
Individual octet at position $j$ & $o_j \in \mathbb{O}$ \\
Transformed octet & $o'_j \in \mathbb{O}$ \\
IP address vector & $\mathbf{ip} = \langle o_1, o_2, o_3, o_4 \rangle$ \\
Anonymized IP address vector & $\mathbf{ip}' = \langle o'_1, o'_2, o'_3, o'_4 \rangle$ \\
Cryptographic salt & $\boldsymbol{\sigma} \in \{0,1\}^*$ \\
SHA-256 hash function & $\mathcal{H} : \{0,1\}^* \rightarrow \{0,1\}^{256}$ \\
Position-specific transformation table & $\mathcal{T}_j : \mathbb{O} \rightarrow \mathbb{O}$ for $j \in \{1,2,3,4\}$ \\
Concatenated hash input & $\xi \in \{0,1\}^*$ \\
Hash digest & $\delta \in \{0,1\}^{256}$ \\
String concatenation operator & $\parallel$ \\
String encoding function & $\text{enc}$ \\
Binary-to-integer conversion & $\text{bin2int}$ \\
Function domain & $\text{dom}(f)$ \\
\end{tabular}
\end{center}
}
\twocolumn


\section{Security Analysis and Demonstration of the Proposed Scheme}
\label{sec:security}
Analysis of security of our salt-based hashing anonymization technique for IP addresses and port numbers and provide an example to illustrate.

\textbf{Salt Generation and Secrecy:}\\ 
A unique random salt is generated per file, which ensures that even if an adversary knows the hash function (e.g., SHA-256), they cannot precompute hash values for each possible octet. It is important to keep this salt a secret as it can be used for brute force if leaked. We recommend using it in memory and then discarding it.

\textbf{Cryptographic Hash Function:} \\
The scheme uses a strong cryptographic hash function such as SHA-256. This function is non-invertible. No preimage attacks possible. Without the salt, recovering the original octet value is computationally infeasible. \\
\textbf{The random file salt along with the cryptographic hash function makes learning based attacks infeasible.} The security foundation of the anonymization scheme rests upon the cryptographic assumptions of SHA-256 security. Under these assumptions, recovering the original IP address $\boldsymbol{\alpha}$ from its anonymized counterpart $\boldsymbol{\alpha}'$ without knowledge of the salt $\boldsymbol{s}$ is computationally infeasible for any polynomially-bounded adversary. This computational anonymity property relies on the preimage resistance of SHA-256, where given $\boldsymbol{\alpha}' = \Phi(\boldsymbol{\alpha})$, an adversary must invert the hash function for each octet position. Without the salt $\boldsymbol{s}$, this requires solving the preimage problem for SHA-256, which is computationally intractable under standard cryptographic assumptions. The cryptographic security of the anonymization scheme is critically dependent on the secrecy and entropy of the salt $\boldsymbol{s}$. Compromise of $\boldsymbol{s}$ enables complete reconstruction of the mapping tables through exhaustive enumeration of all possible octet values and their corresponding hash outputs.
\begin{table}[htbp]
\centering
 \renewcommand{\arraystretch}{1.3}
    \resizebox{\columnwidth}{!}{%
\begin{tabular}{|c|c|c|c|c|}
\hline
\textbf{Run} & \textbf{Avg Collision} & \textbf{Original Entropy} & \textbf{Anonymized Entropy} & \textbf{Entropy Change} \\
\hline
0 & 0.2804 & 11.4157 & 11.4157 & 0.0000 \\
1 & 0.2882 & 11.4157 & 11.4157 & 0.0000 \\
2 & 0.2686 & 11.4157 & 11.4157 & 0.0000 \\
3 & 0.2529 & 11.4157 & 11.4157 & 0.0000 \\
4 & 0.2647 & 11.4157 & 11.4157 & 0.0000 \\
5 & 0.2882 & 11.4157 & 11.4157 & 0.0000 \\
6 & 0.2608 & 11.4157 & 11.4157 & 0.0000 \\
7 & 0.2804 & 11.4157 & 11.4157 & 0.0000 \\
8 & 0.2961 & 11.4157 & 11.4157 & 0.0000 \\
9 & 0.2608 & 11.4157 & 11.4157 & 0.0000 \\
\hline
\end{tabular}
}
\caption{Average collision and entropy metrics across 10 anonymization runs}
\label{tab:anonymization_stats}

\end{table}

\textbf{Domain Reduction and Collision Risk:} \\
Each IP address is divided into four octets. After hashing, the values are mapped into the range 0--255. This constrained range may lead to collisions, it is an acceptable trade-off to preserve the subnet structure. Various ways are possible-can take modulo 256 of the hashed value or map each hash to some random value in 0-255. The same original octet will give the same mapped value within a file, ensuring consistency. \\
 
\begin{table}[htbp]
\centering
\renewcommand{\arraystretch}{1.3}
    \resizebox{\columnwidth}{!}{%
\begin{tabular}{|c|c|c|c|}
\hline
\textbf{Run} & \textbf{Collision Octet 4} & \textbf{Original Entropy} & \textbf{Anonymized Entropy} \\
\hline
0 & 0.2804 & 7.9944 & 7.1373 \\
1 & 0.2098 & 7.9944 & 7.3153 \\
2 & 0.2804 & 7.9944 & 7.1568 \\
3 & 0.2569 & 7.9944 & 7.2009 \\
4 & 0.2647 & 7.9944 & 7.1792 \\
5 & 0.2647 & 7.9944 & 7.1696 \\
6 & 0.2647 & 7.9944 & 7.1989 \\
7 & 0.2569 & 7.9944 & 7.1992 \\
8 & 0.2608 & 7.9944 & 7.2097 \\
9 & 0.2882 & 7.9944 & 7.1210 \\
\hline
\end{tabular}
}
\caption{Collision in Octet 4 vs Entropy Before and After Anonymization}
\label{tab:collision_entropy_summary}
\end{table}

\textbf{The collisions also provide some form of anonymity as each octet is not uniquely mapped to a unique number from 0-255. This makes inference difficult and each run leads to collisions between different octet values making it difficult to predict. The probability of 2 IP's mapping to same IP's after anonymization is very low and hence the shanon entropy remains nearly the same.}
 
\textbf{Determinism and Consistency:} \\
Because the mapping is deterministic (using the file-specific salt), the same IP address always maps to the same anonymized IP within a log file. However, if an attacker were to gain access to multiple files with the same salt or deduce the salt, the deterministic nature could be exploited. So salts need to be unique and confidential per file(or set of files in a computation). It is chosen plain text attack secure.
\paragraph{IND‐CPA Security (Sketch)}  
Let 
\[
E_s(x) \;=\; \bigl(\mathrm{SHA256}(s\parallel x)\bigr)\bmod 256
\]
with secret salt \(s\).  In the random‐oracle/PRF model, SHA‐256 keyed by \(s\) is a pseudorandom function, so \(E_s\) is indistinguishable from a truly random mapping.  Hence no polynomial-time adversary with chosen-plaintext access can distinguish or invert \(E_s\) with non-negligible advantage, i.e.\ it is IND-CPA secure.

Analysis of the anonymization on a column of IP of around 2000 entries was done and recorded in Table~\ref{tab:info_loss_utility}. The IP ranges across a wide range with each octet having 0-255 at least once. The average collision rate calculated as number of octets which get mapped to already mapped anonymized octet by the total original octets (256 in this case). The entropy of original and anonymized remains nearly the same as there's consistency in the anonymization.

Another Table~\ref{tab:collision_entropy_summary} generated has the similar metrics for IP's of the form 192.168.1.i where i goes from 0 to 255. Entropy is bound to decrease as number of uniques IP's is decreasing in the anonymized due to collision but this represents that it is much harder to reverse and gives privacy as well.  

\newpage
\begin{figure}[ht]
    \centering
    \includegraphics[width=0.95\linewidth]{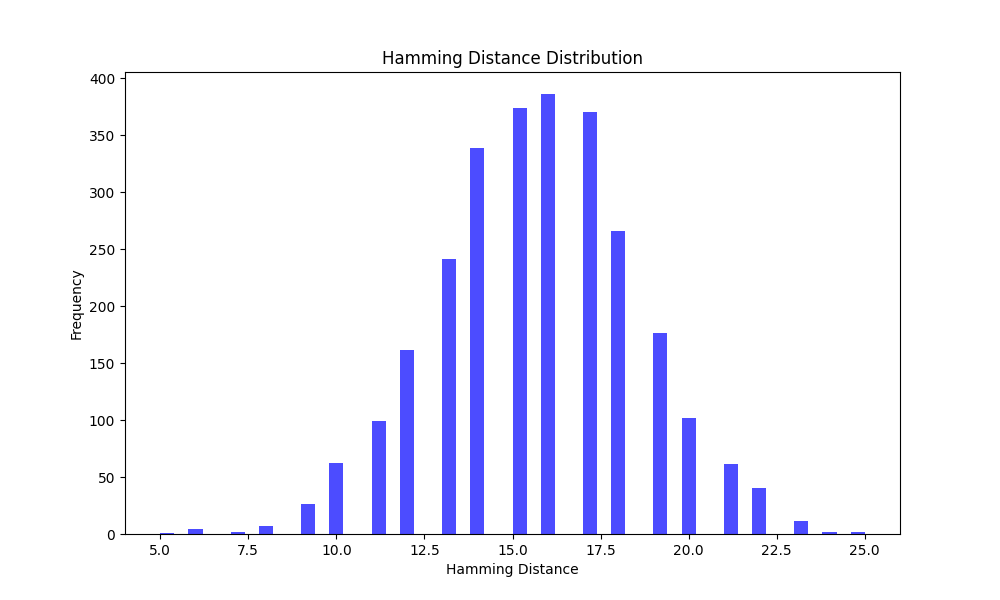}
    \caption{Hamming distance between bitwise representation of IP and its anonymized value}
    \label{fig:IP Anonymisation 2}
\end{figure}

Based on the Hamming distance distribution in Figure~\ref{fig:IP Anonymisation 2}, the anonymization scheme demonstrates effectiveness through its bell-shaped curve centered around 15-16 bits, indicating approximately half the bits are flipped during transformation. This symmetric distribution with low frequencies at extremes shows the method avoids both trivial changes and excessive alterations, achieving optimal balance between privacy protection and data utility.

\section{Tool}
\label{sec:tool}
A python tool has been developed which implements the methods discussed above. The tool offers multiple functionalities with a log parser, an anonymization module and a log reconstructor to fill the anonymized values back in place. There is support for custom parsing as well as fixed format for suricata, zeek, etc. Various anonymization techniques are available for each field. The main objective was to make conversion of logs into anonymized logs easily without much change in the format of the logs to make it easier to feed it for further analysis. The PIIs it can identify are IPs, ports, timestamps and payload data size. It can further be modified easily based on use case. The github link is here. \\

\textit{\url{https://github.com/asdf1879/Network-Log-anonymisation}}

\begin{table}[htbp]
\centering
\caption{Information Loss vs.\ Retained Utility}
\label{tab:info_loss_utility}
\renewcommand{\arraystretch}{1.5}
    \resizebox{\columnwidth}{!}{%
\begin{tabular}{|l|p{5cm}|p{6cm}|}
\hline
\textbf{Field}        & \textbf{Status After Anonymization}                                               & \textbf{Impact on Analysis}                                                                                       \\
\hline
IP Address            & Salt‐based per‐octet hash (mod 256), file‐unique salt, partial prefix preserved   & Subnet‐level grouping preserved; intra‐log correlation still possible; cannot link to real IPs externally        \\
MAC Address           & Consistent one‐way hash (salted)                                                  & Device uniqueness preserved within file; no vendor/OUI info; cross‐file linking broken                           \\
Port Number           & Salt‐based hash $\to$ mod 65536                                                   & Port‐usage distributions intact; service inference still possible; exact port identity hidden                     \\
Timestamp             & Order‐preserving adaptive noise + global offset                                   & Sequence and relative intervals preserved; exact times shifted; peak/trend analysis remains valid                \\
Payload / URL         & Masked/tokenized                                                                  & Behavioral intent (endpoints hit) visible; sensitive parameters removed                                          \\
TCP Seq.\ No.         & Random permutation (consistent within file)                                       & Packet‐ordering and flow analysis possible; exact sequence numbers cannot be abused                              \\
User-Agent            & Generalized to device class (browser/OS family). Use a placeholder or enumeration.                                   & Device fingerprinting reduced to coarse classes; useful for broad client‐type analytics                          \\
\hline
\end{tabular}
}
\end{table}

\begin{table}[htbp]
\centering
\caption{Residual Leakage Analysis (After Anonymization)}
\label{tab:residual_leakage}
\renewcommand{\arraystretch}{1.5}
    \resizebox{\columnwidth}{!}{%
\begin{tabular}{|l|c|p{7cm}|c|}
\hline
\textbf{Field}       & \textbf{Residual Leakage}                                                                                 & \textbf{Risk Level} \\
\hline
IP Address         & Subnet sizes and active‐subnet patterns can be inferred; cross‐file correlation broken if salts differ   & Low    \\
MAC Address                         & Device usage frequency and “unique device” counts still visible within a file                            & Low    \\
Port Number                              & Popular service ports (e.g., 80, 443) may still stand out by frequency                                   & Low    \\
Timestamp                                 & Time‐of‐day patterns (e.g., business hours) remain; inter‐event timing slightly perturbed                 & Medium \\
Payload                         & parameter values hidden but retains the rough value estimate                & Low    \\

User-Agent                               & Broad device class (mobile vs.\ desktop) still identifiable; fine‐grained fingerprinting not possible     & Medium \\
\hline
\end{tabular}
}
\end{table}

\section{Conclusion}
\label{sec:conclusion}
This research tackled how to share network log data securely without losing its value for security analysis. Through extensive experimentation, we created techniques that significantly improve upon existing approaches by treating different data types with specialized protection methods.
Our main contribution was developing IP and port anonymization via salt-based hashing that keeps network relationships intact while making addresses untraceable. Traditional methods either randomize completely, breaking network analysis, or use simple substitution that's easily reversed. Our salt-based hashing processes each IP octet separately with secret keys that change per log file, letting researchers see network relationships and track connections without determining actual addresses or linking across datasets. For timestamps, we solved how to hide exact timing while preserving event order. Previous approaches scrambled sequences, changing security incident meanings. Our adaptive noise technique adjusts perturbation based on neighboring event distances, keeping relative order while preventing chronological determination.
The approach maintains compatibility with existing security tools by preserving log format, removing a major data sharing barrier. Our open-source tool processes data efficiently with modular structure for easy extension.
Testing with real security data showed essential analytical capabilities survive anonymization. Researchers can identify attack patterns, track malicious behavior, and develop detection methods while only losing the ability to identify specific systems.
This work enables better cybersecurity collaboration. When organizations can safely share quality log data, the security community becomes more effective at detecting and stopping attacks. Our research provides practical tools balancing privacy protection with security effectiveness.

\nocite{*} 
\bibliographystyle{IEEEtran}
\bibliography{references}

\newpage
\onecolumn
\appendix
\section{Network Log Anonymization Tool Implementation}
\label{appendix:tool_implementation}

\subsection{Tool Architecture}
The tool consists of modular components: core pipeline (\texttt{main.py}, \texttt{log\_parser.py}, \texttt{log\_reconstructor.py}), anonymization modules (\texttt{ip\_anonymizer.py}, \texttt{port\_anonymizer.py}, \texttt{timestamp\_anonymizer.py}), privacy algorithms (\texttt{differential.py}, \texttt{l-diversity.py}, \texttt{t\_closeness.py}), and configuration (\texttt{config.yaml}). The tool implements two key algorithms proposed in this paper: \textbf{Order-Preserving Timestamp Anonymization with Adaptive Noise} maintains temporal ordering while adding calibrated noise to timestamps, rounding to nearest intervals and applying adaptive noise based on traffic density to preserve privacy while maintaining sequence integrity. \textbf{IP and Port Anonymization via Salt-Based Hashing} uses SHA-256 with configurable salt values for consistent one-to-one mapping, processing IP addresses at octet-level to preserve network structure while mapping ports to non-reserved ranges (1024-65535) maintaining service relationships.

\subsection{Configuration Schema}
\begin{verbatim}
log_file: input_path
log_type: [suricata|firewall|zeek]
output_log: output_path
anonymization:
  ip: {method: salt }
  port: {method: salt}
  timestamp: {method: adaptive}
\end{verbatim}

\subsection{Usage}
Configure parameters in \texttt{config.yaml}, execute 
\begin{verbatim}
python main.py --config config.yaml 
\end{verbatim}

\subsection{Anonymization Results}
The tool processes network logs as demonstrated below:

\textbf{Original Log Sample:}
\begin{verbatim}
03/17/2025-22:48:07.698063  192.168.1.178:57621 -> 192.168.1.255:57621
03/17/2025-22:48:11.711782  192.168.2.184:57621 -> 192.168.1.255:57621
03/17/2025-22:49:19.469764  192.168.1.181:8080 -> 192.168.1.255:8080
03/17/2025-22:49:24.544901  192.168.3.204:3389 -> 192.168.1.255:3389
03/17/2025-22:49:27.132054  203.0.113.76:80 -> 192.168.1.192:36734
03/17/2025-22:49:27.684775  192.168.1.223:22 -> 192.168.1.255:22
03/17/2025-22:49:29.205681  192.168.2.82:443 -> 192.168.1.255:443
03/17/2025-22:49:38.263593  192.168.1.146:57621 -> 192.168.1.255:57621
\end{verbatim}

\textbf{Anonymized Log Output:}
\begin{verbatim}
03/17/2025-21:57:50.402  129.195.79.72:63917 -> 129.195.79.250:63917
03/17/2025-21:57:52.817  129.195.247.221:63917 -> 129.195.79.250:63917
03/17/2025-21:58:59.425  129.195.79.140:54030 -> 129.195.79.250:54030
03/17/2025-21:59:06.576  129.195.164.221:27366 -> 129.195.79.250:27366
03/17/2025-21:59:08.237  63.165.104.225:57270 -> 129.195.79.129:59620
03/17/2025-21:59:08.613  129.195.79.238:40993 -> 129.195.79.250:40993
03/17/2025-21:59:09.989  129.195.247.28:55577 -> 129.195.79.250:55577
03/17/2025-21:59:19.254  129.195.79.124:63917 -> 129.195.79.250:63917
\end{verbatim}

The results demonstrate consistent anonymization with preserved network relationships, temporal ordering, and service port mappings while protecting sensitive information through salt-based hashing and adaptive noise injection.
\end{document}